\magnification = 1200
\voffset 0.5 true in
\baselineskip 20 pt
\nopagenumbers

\centerline {\bf VARIATIONAL APPROACH TO GAUSSIAN APPROXIMATE
COHERENT}

\centerline {\bf STATES: QUANTUM MECHANICS AND MINISUPERSPACE}

\centerline {\bf FIELD THEORY}
\vskip 0.5 true in

\centerline {A. A. Minzoni}

\centerline {\it FENOMEC/Department of Mathematics and Mechanics, IIMAS\/}

\centerline {\it Universidad Nacional Aut\'onoma de M\'exico\/}

\centerline {\it A. Postal 20-726\/}

\centerline {\it M\'exico 01000 D.F., MEXICO\/}
\vskip 10 pt

\centerline {Marcos Rosenbaum and Michael P. Ryan, Jr.}

\centerline {\it Instituto de Ciencias Nucleares\/}

\centerline {\it Universidad Nacional Aut\'onoma de M\'exico\/}

\centerline {\it A. Postal 70-546\/}

\centerline {\it M\'exico 04510 D.F., MEXICO\/}

\vfill\eject
\voffset 0 in

\noindent {\bf ABSTRACT}
\vskip 10 pt

This paper has a dual purpose.  One aim is to study the evolution of 
coherent states in ordinary quantum mechanics.  This is done by means of
a Hamiltonian approach to the evolution of the parameters that define 
the state.  The stability of the solutions is studied.  The second aim is 
to apply these techniques to the study of the stability of minisuperspace
solutions in field theory.  For a $\lambda \varphi^4$ theory we show, both by
means of perturbation theory and rigorously, by means of theorems of the
K.A.M. type, that the homogeneous minisuperspace sector is indeed stable
for positive values of the parameters that define the field theory.
\vfill\eject

\pageno = 1
\footline{\hss\tenrm\folio\hss}
\noindent {\bf I. INTRODUCTION}
\vskip 10 pt

This paper has a dual purpose.  It has its origin in the study of
superspaces in field theory, that is, the function spaces of
solutions of any field theory (even though the name 
superspace originated in the
study of the gravitational field).  However, we would like to
emphasize that the techniques we will use here may be applied to the study
of coherent states in ordinary quantum mechanics, and the examples we
will give are, in fact, equivalent to one- and two-dimensional
nonrelativistic quantum mechanics.

The simplest example of a superspace is that of a one-dimensional
real scalar field $\varphi (z, t)$.  If we expand $\varphi$ in a real
Fourier series (assuming the domain of 
$\varphi$ to be confined to $-L/2 < z < L/2$
with the end points identified)
$$\varphi (z, t) = \varphi_0 (t) + \sum_{n = 1}^{\infty} \left \{ \varphi_n
(t) \cos \left ( {{2n\pi z}\over {L}} \right ) + \varphi_{-n} (t) \sin
\left ( {{2n\pi z}\over {L}} \right ) \right \}, \eqno (1.1)$$
the evolution of $\varphi$ (independent of the action that generates
the evolution) is nothing more than a curve in the space of
countably infinite dimension defined by the ``coordinates''
$\varphi_n$, $-\infty \leq n \leq +\infty$.  Of course, classically any
nonlinear action for $\varphi$ gives an extremely complicated infinite set of
coupled ODE's for the $\varphi_n$, and this approach to field theory
is rarely used in direct calculations.

Nevertheless, such Fourier expansions were the basis for field
quantization for many years, and are still used in many contexts.  It
is not difficult to show that for a real Fourier series such as (1.1)
the quantum evolution of $\varphi$ is just the quantum mechanics of a
particle moving in $\varphi_n$-space under the influence of what may
be a very
complicated potential (even in the simplest cases of nonlinear
actions for $\varphi$).  There have been a number of studies of
nonlinear field theories which have attempted to glean information
about the behavior of fields such as $\varphi (z, t)$ by studying
model theories where the configuration space of the system is reduced
by putting all but a finite number of the $\varphi_n$ equal to zero
[1, 2, 3, 4].
These are called ``minisuperspace'' field theories.  Notice that such
theories are equivalent to one-particle quantum mechanics in a space
of dimension of the surviving $\varphi_n$.  One may ask whether such
theories are just models or whether they could be approximate
solutions to the full field theory.  One method for answering this
question is to study approximate ``coherent states'' whose center is
supposed to move on a track in $\varphi_n$-space that is centered on
some classical path which is restricted to the reduced configuration
space.

This problem leads to the study of the simpler quantum mechanical
problem of finding consistent approximations
to the time-dependent evolution of localized solutions to the Schr\"odinger
equation in dimensions corresponding to the reduced configuration
space.  Of course, a problem of this sort is completely independent
of any field theory, and we feel that our results are useful in the
general study of coherent states.

An interesting method for the study of these states is based on the ideas of 
average variational principles.  This approach is widely used in the context
of nonlinear waves and oscillations [5] and more recently in the
problem of soliton propagation under the influence of various perturbing
effects [6].  The
emphasis here will be on obtaining approximate and rigorous stability
results rather than on detailed calculations of the evolution of the
solution which is the purpose of many works on wave propagation. 
These stability arguments are of much more interest for field theory,
since what has been called the ``quantum stability'' of
minisuperspace solutions is directly related to the problem of the
use of quantum minisuperspace solutions as approximations, but we will
find evolution equations for coherent state solutions which could be
of use in the study of coherent states in ordinary quantum
mechanics.

The technique we will use is based on the consideration of
time-dependent trial functions in the
Lagrangian of the Schr\"odinger equation.  The Lagrangian is then 
averaged over the space variables to obtain an effective action which
involves only the time-dependent parameters of the trial function.  The
Euler equations of this new Lagrangian give us the evolution of the 
parameters.  This procedure gives a consistent way to approximate 
the infinitely many degrees of freedom of the wave function 
by means of a finite number of
parameters while preserving the Lagrangian structure.  On the other hand,
pointwise approximations could, in principle (and often in practice),
produce in the truncation spurious non-conservative terms.  Since we
are interested in a Hamiltonian formulation of the problem, its stability
is not determined by damping terms but rather by the nature of of the
Hamiltonian.  Because of this the approximation must be consistent with
the underlying Hamiltonian structure.  We would like to emphasize once
more that this idea gives us a way to construct approximate coherent
states in quantum mechanics and may be very useful in what has become
a very active field.

In Ref. [4] a pointwise approximation to the motion of ``Gaussian''
wave packets in minisuperspaces of a $\lambda \varphi^4$ theory was
presented where the action was
$$L = \int \left [ {{1}\over {2}} \left ( {{\partial \varphi}\over
{\partial t}}\right )^2 - {{1}\over {2}}\left ( {{\partial
\varphi}\over {\partial z}} \right )^2 -{{\mu^2}\over {2}} \varphi^2 -
\lambda \varphi^4 \right ] dz dt, \eqno (1.2)$$
and an $S^1$ topology was assumed for $t =$ const. slices,
identifiying the end points $z = \pm L/2$.  The field
$\varphi$ was then expanded in the same Fourier series as (1.1).   
One- and two-dimensional minisuperspaces were
studied, with special attention to the quantum stability of solutions
of the one-dimensional sector as imbedded in the two-dimensional
sector. The one-dimensional superspace had a state
function of the form $\Psi (x, t)$, where $x$ was equal to
$\varphi_0$ of (1.1).  This function obeyed a non-relativistic
Schr\"odinger equation of the form
$$i{{\partial \Psi}\over {\partial t}} = -{{1}\over {2}} {{\partial^2
\Psi}\over {\partial x^2}} + \left [ {{1}\over {2}} \mu_0^2 x^2 +
\varepsilon x^4 \right ] \Psi, \eqno (1.3)$$
with $\varepsilon = \lambda /L$.  Notice that this is nothing more
than a one-particle Schr\"odinger equation for a particular
anharmonic potential.

The behavior of localized packets of the form $\Psi = e^{-{\cal S}}$
was then considered, where ${\cal S}$ was taken to be
$${\cal S} = \alpha (t) x^4 + \beta (t) x^3 + \gamma (t) x^2 + \sigma
(t) x^2 +$$
$$ {{\mu_0}\over {2}} [x - g(t)]^2 + i B(t) x^3 + iC(t) x^2 + i[P(t)
+ D(t)] x. \eqno (1.4)$$
To derive the evolution of ${\cal S}$ a pointwise
approximation resulting from inserting
$\Psi = e^{-{\cal S}}$ in (1.3) was used.  The
parameters $\alpha$, $\beta$, $\gamma$, $\sigma$, $B$, $C$, $D$ were
assumed to be of order $\varepsilon$, and a consistent set of coupled linear
ODE's for these parameters was obtained.

The two-dimensional minisuperspace was one where $\varphi_0$ and {\it
one \/} of the $\varphi_n$ were taken to be non-zero.  This will be
discussed in more detail in Section IV.

The equations for the parameters in both the one- and two-dimensional
minisuperspaces
could be solved recursively, but the theory
predicted secular growth of both the ``center of mass'' of the state
and the localization parameter (controling ``spreading'').  In this
paper we will use the method outlined above 
to show that the motions are modulated and,
hence, stable. Since the
motion is Hamiltonian, it is not possible to have asymptotic
stability,
and the averaging results are not conclusive for all time.  However,
it is possible to show rigorously, by applying the 
Kolmogorov, Arnold, Moser (K.A.M.) theorem [7], that
special motions are indeed stable for all time.

The paper is organized as follows.  In Section II we study the
approximate evolution of localized states in ordinary quantum
mechanics by means of a variational approximation where we use (1.3)
as an example.  In Section III we present perturbation equations for
the parameters of the localized state of Sec. II and show that their
solutions are bounded, and then use the K.A.M. theorem to show
stability for the full equations.  In Section IV we study the
two-dimensional minisuperspace field theory using a theorem by
Melnikov, Poesch, and Kuskin [8] to show that the solutions are stable
for a set of full measure in the parameter space. Section V contains
conclusions and suggestions for further research.  In Appendix A we
present the equations for a pointwise approximation for $\Psi =
e^{-{\cal S}}$, using an ${\cal S}$ similar to (1.4).  Appendix B
contains the full action for the parameters of the two-dimensional
minisuperspace solutions.
\vfill\eject
   
\noindent {\bf II. VARIATIONAL APPROXIMATION}

In this section we will study the behavior of an approximate coherent
state in ordinary quantum mehanics for a particle moving in the
potential $V(x) =(\mu_0^2/2) x^2 + \varepsilon x^4$, where
$\varepsilon = \lambda /L$ gives us the one-dimensional minisuperspace
theory.  Since $L$ will be something like a ``radius of the
universe'' in the minisuperspace case, we will assume that it is very
large, and thus concern ourselves with potentials where $\varepsilon$
is small.  In order to find the behavior of our state, we will appeal
to various approximations, but
of course there is nothing to say that an
exact ``coherent state'' does not exist for this $V(x)$.  In fact,
depending on the definition one takes for a coherent state, one
should be able to find such exact solutions.

For example, if we take a one-dimensional system moving in an
arbitrary potential $\tilde V (x)$, and assume that a solution of the
form $\Psi = e^{-{\cal S} (x,t)}$ exists, where ${\cal S} = {\cal S} [x,
x- w(t)]$, and define $u \equiv x$, $v \equiv x - w(t)$, ${\cal S}$
will satisfy the nonlinear partial differential equation
$$ i\dot w [w^{-1} (u - v)] {{\partial {\cal S}}\over {\partial v}} +
{{1}\over {2}} {{\partial^2 {\cal S}}\over {\partial u^2}} +
{{\partial^2 {\cal S}}\over {\partial u \partial v}} + {{1}\over {2}}
{{\partial^2 {\cal S}}\over {\partial v^2}} -$$
$$-{{1}\over {2}}\left ({{\partial {\cal S}}\over {\partial u}}\right
)^2 - {{\partial {\cal S}}\over {\partial u}} {{\partial {\cal
S}}\over {\partial v}} - {{1}\over {2}} \left ( {{\partial {\cal
S}}\over {\partial v}}\right )^2 + \tilde V (u) = 0. \eqno (2.1)$$
The boundary conditions must be taken such that ${\cal S} \rightarrow
\infty$ as $x \rightarrow \pm \infty$.  The maximum value of $\Psi$
will occur where $(\partial {\cal S}/\partial u ) \pm (\partial {\cal
S}/ \partial v ) = 0$, which can be solved for $x = F(t)$.  The
function $w$ can be chosen to make the peak value follow the
classical motion of a particle moving in $\tilde V(x)$.  There should
be no a priori reason why (2.1) would be impossible to solve for these
boundary conditions for many potentials $\tilde V(x)$.  However, for
most $\tilde V(u)$ an analytic solution for ${\cal S}$ will be
difficult to find, and most solutions for coherent states are only
approximate.

The pointwise approximation mentioned in the Introduction consisted
in using $\Psi = e^{-{\cal S}}$, with ${\cal S}$ from (1.4), in
(1.3).
However, there are a number of other ways to approximate such a
state.  One that is frequently used in the study of solitonic
solutions to nonlinear differential equations is to insert a simple
Gaussian trial function with time-dependent parameters into the
action for the theory.  In our case we can do the same, inserting
such a function into the usual action for the Schr\"odinger equation
$$ L = \int^{t_1}_{t_0} dt \int^{+\infty}_{-\infty} dx [i\Psi^{*}
\Psi_t - {{1}\over {2}} \Psi^{*}_x \Psi_x - V(x) \Psi^{*} \Psi ].
\eqno (2.2)$$
If we take a Gaussian ansatz for $\Psi$, $\Psi \equiv W e^{-\cal S}$,
where $W = W(t)$ and
$$ {\cal S} = {{\mu(t)}\over {2}} [x - g(t)]^2 + i P(t)x + iC(t)x^2 +
i\phi (t), \eqno (2.3)$$
the $g(t)$ gives the center of the probability packet $\Psi^{*}
\Psi$, and $\mu (t)$ allows for ``breathing'' or ``spreading'' of the
packet.  Plugging this expansion into (2.2) and doing the
$x$-integrations, we find
$$I = \sqrt{\pi}\int^{t_1}_{t_0} {{W^2}\over {\sqrt {\mu}}} \bigg \{ \dot P
g + \dot C \left ( g^2 + {{1}\over {2\mu}} \right ) + \dot \phi - \bigg [
{{P^2}\over {2}} + {{\mu_0^2 g^2}\over {2}} + \varepsilon g^4 +$$ 
$$ + {{3\varepsilon}\over {\mu}} g^2 + 2C^2 \left ( g^2 + {{1}\over
{2\mu}} \right ) + 2PCg + {{\mu}\over {4}}\left ( 1 + {{\mu_0^2}\over
{\mu^2}} \right ) + {{3}\over {4}} {{\varepsilon}\over {\mu^2}}
\bigg] \bigg \} dt \eqno (2.4)$$
In the ${\cal S}$ given in (1.4) one had to assume that many of the
parameters were of order $\varepsilon$ so that a consistent
set of equations could be found.  Here we
are no longer forced to suppose that the deviations of these
quantities from their harmonic-oscillator values are small, and we
will see that a much simpler set of equations emerge.

The Euler equations for (2.4) are
$$\left ( {{W^2}\over {\sqrt{\mu}}} \right )^{\bf \cdot} = 0, \eqno
(2.5a)$$ 
$$-\dot g  - P - 2Cg = 0, \eqno (2.5b)$$
$$\dot P - \mu_0^2 g - 4\varepsilon g^3 - {{6\varepsilon}\over {\mu}}
g + 2\dot C g - 4 C^2 g - 2PC = 0, \eqno (2.5c)$$
$$-2 g \dot g + {{\dot \mu}\over {2\mu^2}} - 4C\left ( g^2 +
{{1}\over {2\mu}} \right ) - 2Pg = 0, \eqno (2.5d)$$
$${{\dot C}\over {2\mu^2}} + {{C^2}\over {\mu^2}} -
{{3\varepsilon}\over {\mu^2}}g^2 - {{1}\over {4}} + {{\mu_0^2}\over
{4\mu^2}} - {{3}\over {2}} {{\varepsilon}\over {\mu^3}} = 0, \eqno
(2.5e)$$
together with an equation which results from varying $W$ 
that gives $\phi$ as a quadrature
once $g$, $P$, $\mu$, and $C$ are known.  Equations (2.5) can be
solved recursively.  Using (2.5b) and (2.5d) we obtain
$$C = {{\dot \mu}\over {4\mu}}. \eqno (2.6)$$
From (2.5c) and (2.5e) we find the equation of motion for the center
of the packet,
$$\ddot g + \mu_0^2 g + 4\varepsilon g^3 + {{6\varepsilon}\over
{\mu}} g = 0. \eqno (2.7)$$
Finally, using (2.6) in (2.5e) we find
$$-{{\ddot \mu}\over {\mu}} + {{3}\over {2}} \left ( {{\dot \mu}\over
{\mu}} \right )^2 - 2\mu^2 + 2\mu_0^2 - {{12\varepsilon}\over {\mu}}
+ 24 \varepsilon g^2 = 0. \eqno (2.8)$$
If we introduce the change of variables $\mu \equiv 1/a^2$, we find
the final system
$$\ddot g + \mu_0^2 g + 4\varepsilon g^3 + 6\varepsilon a^2 g = 0,
\eqno (2.9a)$$
$$\ddot a + \mu_0^2 a - {{1}\over {a^3}} - 12 \varepsilon a^3 + 24
\varepsilon a g^2 = 0, \eqno (2.9b)$$
for the position $g$ and the variance (``breathing'') of our
approximate coherent state.

In order to compare these equations for $\mu$ and $g$ with those in
Ref. [4], we can see that if we write, as above, $\mu = \mu_0 +
2\varepsilon \gamma$, and $g = g_0 + 2\varepsilon \sigma$ and keep
terms to order $\varepsilon$, we find
$$\ddot g_0 + \mu_0^2 g_0 = 0, \eqno (2.10a)$$
$$\ddot \sigma + \mu_0^2 \sigma = -2g_0^3 - {{3g_0}\over {\mu_0}},
\eqno (2.10b)$$
and
$$\ddot \gamma + 4\mu_0^2 \gamma = 6 + 12 g_0^2. \eqno (2.11)$$
Instead of comparing these equations directly with those of Ref. [4],
it is more useful to use the parametrization used in this article and
apply it to the pointwise approximation of Ref [4].  To do this we
write
$${\cal S} = \alpha (t) x^4 + \beta (t) x^3 + {{\mu (t)}\over {2}} [x
- g(t)]^2 + iB(t)x^3 + iC(t)x^2 + iP(t)x + i\phi (t). \eqno (2.12)$$
(leaving out a trivial real $x$-independent term which appears in Ref.
[4]).  We have to assume that $\alpha$, $\beta$, and $B$ are small
enough that we only keep terms to linear order in them or we will
have terms in $x$ of order higher that $x^4$, and the ansatz will be
inconsistent.  We can substitute this ${\cal S}$ in the Schr\"odinger
equation (1.3) with $\Psi = e^{-{\cal S}}$ and find a series of
coupled equations for $\alpha$, $\beta$, $\mu$, $g$, $B$, $C$, and
$P$ (we will ignore the equation for $\phi$).  These are given in
Appendix A.  The equations for $\alpha$, $\beta$, and $B$ are the
same as in Ref. [4].  If we take $\mu = \mu_0 + 2\varepsilon \gamma$ and
$g = g_0 + 2\varepsilon \sigma$, and assume that $\alpha$, $\beta$,
and $B$ are of order $\varepsilon$, we find $\dot \alpha = 0$,
$\alpha = \varepsilon/4\mu_0$.  Equations (A3) and (A4) imply
$$\ddot \beta + 9\mu_0^2 \beta - 4\varepsilon \mu_0 g_0 = 0, \eqno
(2.13)$$
$$B = {{\dot \beta}\over {3\mu_0}} - {{\varepsilon}\over {3\mu_0^2}}
+ {{\varepsilon \dot g_0}\over {3\mu_0^2}}, \eqno (2.14)$$
and the solutions are (taking $g_0 = Q \cos \mu_0 t$)
$$\beta = \varepsilon x_3 \cos 3\mu_0 t + {{\varepsilon Q}\over
{2\mu_0}}\cos \mu_0 t, \eqno (2.15)$$
$$B = \varepsilon x_3 \sin 3\mu_0 t - {{Q\varepsilon}\over {2\mu_0}}
\sin \mu_0 t - {{\varepsilon}\over {3\mu_0^2}}, \eqno (2.16)$$
where $x_3$ is a constant. Now, Eqs. (A6) 
and (A8) can be solved for $C$ and $P$ respectively.
Taking the derivative of the equation for $C$ and inserting the
result in (A5) gives
$$-{{\ddot \mu}\over {\mu}} + {{3}\over {2}} \left ( {{\dot \mu}\over
{\mu}} \right )^2 - 2\mu^2 + 2\mu_0^2 + \varepsilon \bigg ( -6g^2 +
{{6}\over {\mu}} + {{6}\over {\mu^2}} \dot g^2 +$$
$$ + {{6\dot \mu^2}\over {\mu^4}} g^2 + 12 {{\dot \mu}\over {\mu^3}}g
\dot g \bigg ) + 36\mu g \beta + B\left ( -36 \dot g - 33 {{\dot
\mu}\over {\mu}} g \right ) = 0. \eqno (2.17)$$
Substituting the result for $\dot P$ in (A7) gives
$$\ddot g + \mu_0^2 g + \varepsilon \left (6g^3 + {{6\dot g^2 g}\over
{\mu^2}} + {{6\dot \mu^2}\over {\mu^4}} \dot g g \right ) +$$
$$+ \beta \left ( 30\mu g^2 + 12 - {{9}\over {2}} {{\dot \mu^2}\over
{\mu^3}} g^2 - {{21}\over {2}} {{\dot \mu}\over {\mu^2}} g\dot g -
{{6\dot g^2}\over {\mu}} \right ) +$$
$$+ B\left (-36 \dot g g - {{63}\over {2}} {{\dot \mu}\over {\mu}} g^2
\right ) = 0. \eqno (2.18)$$
These two equations can be compared to (2.7) and (2.8).  Note that
they are the same except for differences in the order-$\varepsilon$
``driving terms'' that modify the second order ODE's that determine
$g$ and $\mu$.

In order to compare these equations with (2.10b) and (2.11), we will
again take $\mu = \mu_0 + 2\varepsilon \gamma$, $g = g_0 +
2\varepsilon \sigma$.  To order $\varepsilon$ we again get (2.10a) for
$g_0$ and
$$\ddot \sigma + \mu_0^2 \sigma + \left ( 3g_0^3 + {{3\dot g_0^2
g_0}\over {\mu_0^2}} \right ) + \left ( {{\beta}\over
{\varepsilon}}\right ) \bigg ( 15 \mu_0 g_0^2 + 12 -$$
$$- {{6\dot g_0^2}\over {\mu_0}} \bigg )
-18 \left ({{B}\over {\varepsilon}}\right ) \dot g_0 g_0 = 0, \eqno
(2.19)$$
$$\ddot \gamma - 4\mu_0^2 \gamma + \left ( 3\mu_0 g_0^2 + 3 + {{3\dot
g_0^2}\over {\mu_0}} \right ) + 18 \left ( {{\beta}\over
{\varepsilon}}\right ) \mu_0^2 g_0 - 18 \left ( {{B}\over
{\varepsilon}} \right ) \mu_0 \dot g_0 = 0. \eqno (2.20)$$
Once more these differ from (2.10b) and (2.11) in the ``driving terms'',
where the term proportional to $\varepsilon$ is slightly different
and there are added terms in $B$ and $\beta$.

Clearly, the pointwise approximation, aside from being more complicated, does
not add much to the simpler equations from the variational
approximation.  In Ref. [4] the main problem with the equations for
$\gamma$ and $\sigma$ (i.e. $\mu$ and $g$) is that they generated
terms with secular growth, which made the predictions of the theory
only valid for a limited time.  Both (2.17) and (2.18) have terms
that drive $\sigma$ and $\gamma$ at resonance, so to
order $\varepsilon$ in perturbation they will also have secular terms. 
In Ref. [4] it was assumed that better perturbation methods would
remove such terms.  In the next section we will apply such methods to
the variational approximation as well as provide a rigorous
demonstration that the solutions for $\mu$ and $g$ are indeed
bounded.
\vfill\eject

\noindent {\bf III. ABSENCE OF SECULAR GROWTH}
\vskip 10 pt

\noindent {\bf A. PERTURBATION SOLUTION}

In the previous section we constructed the action and obtained the
equations for $\mu$ (i.e. $a = 1/\sqrt{\mu}$) and $g$.  From (2.10b)
and (2.11) one can see that the order-$\varepsilon$ perturbation
theory leads to secular growth for both the $g$ and $a$
perturbations.  However, there are more sophisticated perturbation
methods methods which remove these terms and provide a uniform 
approximation to the solution.

If we look at Eqs. (2.9) and make certain assumptions about the size
of $a$ and $g$, we can readily show that $g$ must be bounded.  If in the
last term on the LHS of (2.9a) we take $a$ to be little enough
different from $1/\sqrt{\mu_0}$ so that $\varepsilon a^2 \approx
\varepsilon /\mu_0$, (2.9a) then becomes (with no assumption about
the size of $g$)
$$\ddot g = -\left ( \mu_0^2 + {{6\varepsilon}\over {\mu_0}} \right )
g - 4\varepsilon g^3. \eqno (3.1)$$
This is the equation of a particle moving in the potential
$$V(g) = \left ( {{\mu_0^2}\over {2}} + {{3\varepsilon}\over {\mu_0}}
\right ) g^2 + \varepsilon g^4, \eqno (3.2)$$
and $g$ can be solved for in terms of elliptic functions.  The form
of $V$ shows that the motion of $g$ is always bounded.  Here, of
course, we are assuming that $a$ is bounded (and small enough)
so that there are no
large excursions of the $3\varepsilon g^2/\mu_0$ term in the
potential.

In order to show that $a$ is bounded we can take $a$ to be
$(1/\sqrt{\mu_0}) + \tilde a$, where $\tilde a$ is small compared to
$1/\sqrt{\mu_0}$ (but not necessarily of order $\varepsilon$).  As
long as $\varepsilon \tilde a$ is small enough to be ignored, we need
only take $g_0$ in the last term of Eq. (2.9b).   
Taking $g_0(t) = Q \cos \mu_0 t$ we have that the equation for $\tilde
a$ is, to second order,
$$\ddot {\tilde a} + 4\mu^2_0 \tilde a - 6 \mu_0^{5/2} \tilde a^2 =
-12\left ({{1 + Q^2 \mu_0}\over {\mu_0^{3/2}}}\right )\varepsilon - 12 
{{\varepsilon}\over {\mu^{1/2}_0}} Q^2 \cos 2\mu_0 t. \eqno (3.3)$$

To remove the constant forcing to order $\varepsilon^2$ we take $\tilde
a = \rho - {{3(1 + Q^2 \mu_0)}\over {\mu_0^{7/2}}}\varepsilon$ 
and obtain the equation
$$\ddot \rho + (4\mu^2_0 - \varepsilon \kappa \mu_1) \rho - \kappa
\rho^2 = 
\varepsilon F \cos 2\mu_0 t, \eqno (3.4)
$$
where $\kappa = 6\mu_0^{5/2}$, $\mu_1 = 
6\left ( {{1 + Q^2 \mu_0}\over {\mu_0^{3/2}}} \right )$, 
$F = -12 {{Q^2}\over
{\mu_0^{1/2}}}$.  This equation, for $\kappa = 0$, is the equation
for an oscillator forced at resonance.  The equation will now be solved
to show that the nonlinear terms prevent the growth of the solution.  To
obtain the approximate explicit solution we will use the method of
averaging for the Lagrangian
$$L = \int_{t_0}^{t_1} \left \{ -{{1}\over {2}}\dot \rho^2 + {{(4\mu_0^2 +
\varepsilon\mu_1)}\over {2}} \rho^2 - {{\kappa}\over {3}} \rho^3 - 
\varepsilon
F \rho \cos 2\mu_0 t \right \} dt, \eqno (3.5)$$
and use the trial function
$$\rho = \varepsilon^{1/2} [A(\varepsilon t) \cos2\mu_0 t + B(\varepsilon t)
\sin 2\mu_0 t ] + \varepsilon D(\varepsilon t), \eqno (3.6)$$
where the mean value $D$ has to be introduced due to the quadratic nature
of the nonlinearity.  Averaging over the fast time $2\mu_0 t$, we obtain
$$ \bar L = \int_{t_0}^{t_1} 
\left [\varepsilon^2 \left \{ \mu_0 (A^{\prime}B
- AB^{\prime}) + {{\mu_1}\over {4}}(A^2 + B^2) + 
D^2 - {{\kappa}\over
{2}} D(A^2 + B^2) \right \} - {{1}\over {2}}
\varepsilon^{3/2} FA \right ]dt. \eqno (3.6)$$
We now assume the so far arbitrary (but small) amplitude $F$ to be 
$F = \varepsilon^{1/2} F_0$ to balance the resonant forcing and the 
nonlinear effect.  With this, $\varepsilon^{3/2}F \approx {\cal O}
(\varepsilon^2)$ becomes of the same order as the modulated terms (the
argument can, of course, be reversed, assuming that $F \approx {\cal O}(1)$
and rescaling $\rho$ and the time appropriately).

Variation with respect to $D$ gives the usual algebraic equation for the
mean value as 
$$2D - {{\kappa}\over {2}} (A^2 + B^2) = 0, \eqno (3.7)$$
i.e.
$$D = {{\kappa}\over {4}}(A^2 + B^2). \eqno (3.8)$$
Using this result, and integrating by parts, 
we obtain the final form of $\bar L$ as
$$\bar L = 2\varepsilon^2 \mu_0 \int^{t_1}_{t_0} \left [ AB^{\prime} -
\left ( {{\mu_1}\over {8\mu_0}} (A^2 + B^2) + {{\kappa^2}\over {32\mu_0}}
(A^2 + B^2)^2 + {{F}\over {4\mu_0}} A \right )\right ] dt. \eqno
(3.9)$$
Clearly the solutions of the Euler equations are the level curves of the
Hamiltonian
$$H = {{\mu_1}\over {8\mu_0}}(A^2 + B^2) + {{\kappa^2}\over {32\mu_0}}
(A^2 + B^2)^2 + {{F}\over {4\mu_0}}A. \eqno (3.10)$$
It is easy to see that the critical points of the system are the critical
points of $H$.  They satisfy $\partial H/\partial A$ = $\partial H/\partial
B = 0$.  the local minima are centers and the local maxima are saddles.
Also for large $A^2 + B^2$ all curves are closed.  It is then clear that 
all the motions are bounded.  This shows that the resonant forcing is
balanced by the the nonlinearity, giving a bounded motion for $\rho$.  This
in turn implies that the coherent state does not spread indefinitely
away from its original spreading $(\mu_0)^{-1/2}$.  To complete the 
description of the motion we now examine the central points.  They
satisfy 
$${{\partial H}\over {\partial A}} = 0 = {{F}\over {4\mu_0}} +
{{\mu_1 A}\over {4\mu_0}} + {{\kappa^2}\over {8\mu_0}} 
 A (A^2 + B^2),\eqno (3.11a)$$
$${{\partial H}\over {\partial B}} = 0 = B\left [
\left ( {{\mu_1}\over {4\mu_0}}
\right )+ {{\kappa^2}\over {8\mu_0}}(A^2 + B^2)\right ]. \eqno
(3.11b)$$
The only solution for the second equation is $B = 0$, since $\mu_1$
is always greater than zero. When $B = 0$ the
equation for $A$ is the cubic
$$F + \mu_1 A + 
{{\kappa^2}\over {2}} A^3 = 0.
\eqno (3.12)$$
There is only one critical point $(A^{*}, 0)$
where $A^{*}$ is the only solution of (3.12).  Clearly this is a minimum of
$H$.  This represents a periodic solution of finite amplitude $A^{*}$
with period $2\mu_0$.  In this case all level lines of $H$ are
concentric curves.  They give periodic solutions for $A$ 
and $B$, which in turn represent
quasiperiodic motions of $\rho$.  

This perturbation solution shows that small perturbations remain small.
Interestingly enough some of these perturbation results can be proved
rigorously as we will now show.
\vskip 10 pt

\noindent {\bf B. RIGOROUS RESULTS}
\vskip 10 pt

We now establish the boundedness of the motion for not necessarily small
initial conditions of the amplitude $g$ of these oscillations.  We
consider again the Eq. (2.9b).  Here we will assume that 
$g_0 = A\cos \mu_0 t$, where $A$
is not small.  Moreover, assume the initial conditions for $a$ to be
arbitrary, that is, not necessarily close to $1/\sqrt{\mu_0}$.  In this
case we will show that for sufficiently small $\varepsilon$ the solution
for $a(t)$ is always bounded.  To do this we appeal to the classical
results of periodic perturbations of Hamiltonian systems with one degree
of freedom [7, 9]. This pertubation theory is 
based on the Poincar\'e map.  Since the equation is 
periodic with period $2\pi/2\mu_0$ this map is defined as follows. 
Consider the map:
$$[a(0), \dot a(0)] \rightarrow [a(2\pi/2\mu_0), \dot a(2\pi/2\mu_0)],$$
where $(a, \dot a)$ is the solution of (2.9b).  Clearly this map is 
area-preserving since the system is Hamiltonian.  An invariant curve of
this map represents a two-dimensional invariant torus in the space
$(a, \dot a, t)$; $(t\, \, {\rm mod} \, \, 2\pi/2\mu_0)$.  Since two-dimensional 
tori divide three-dimensional space, we have that initial values inside
an invariant torus remain there.  This bounds the solutions.  Thus the
problem of boundedness of solutions for all time is tranformed into the
problem of finding invariant curves for the Poincar\'e map of (2.9b).

The existence of invariant curves is proved using action-angle variables
for Eq. (2.9b).  To study this map 
it is convenient to first introduce action-angle
variables for the unperturbed oscillator.  The transformation is not 
expressed in terms of elementary functions, but an explicit representation
is not necessary in order to derive the results.

The generating function for the canonical formulation is given by
$$W(a, E) = \int_0^a \sqrt{E - \left ( {{\mu_0^2}\over {2}} \xi^2 + 
{{1}\over {2\xi^2}}\right )} d\xi, \eqno (3.13)$$
where the variable $E$ is expressed in terms of the action (the area
enclosed by the orbit)
$$I(E) = 2\int_{a_1(E)}^{a_2(E)}\left \{E - \left({{\mu_0^2}\over {2}}\xi^2
+ {{1}\over {2\xi^2}} \right ) \right \}^{1/2} d\xi. \eqno (3.14)$$
Clearly $I^{\prime} (E) \geq 0$ and $E$ can be given as $E(I)$.  The
generating function becomes $W (a, I)$. We now have
$$p = \dot a = {{\partial W}\over {\partial a}}, \qquad \dot \theta =
{{\partial W}\over {\partial I}} \eqno (3.15)$$
in the new canonical variables $I$ and $\theta$.  In these new variables
the Hamiltonian is independent of $\theta$.

The perturbed problem in these variables takes the form 
$$\dot I = - \varepsilon {{\partial H^{(1)}}\over {\partial \theta}} (\theta,
I, 2\mu_0 t), \eqno (3.16a)$$
$$\dot \theta = {{\partial W(I)}\over {\partial I}}
+ \varepsilon {{\partial H^{(1)}}\over 
{\partial I}} (\theta, I, 2\mu_0 t) \eqno (3.16b)$$
The Hamiltonian $H^{(1)}$ is the image of $(a^2/12)g^2$ under the 
action-angle change of variables.  The explicit form is not needed, just
note that $H^{(1)}$ is a smooth function of the new variables.  The
Poincare map in these variables, if $(I_0, \theta_0)$ is the inital value
and $(I^{(1)}, \theta^{(1)})$ the final value, takes the form
$$I^{(1)} = I_0 - \varepsilon \int^{2\pi/2\mu_0}_0 {{\partial H^{(1)}}\over
{\partial \theta}}[\theta(\xi, I_0, \theta_0), I(\xi, I_0, \theta_0),
2\mu_0 \xi]\, d\xi, \eqno (3.17a)$$
$$\theta^{(1)} = \theta_0 + \int^{2\pi/2\mu_0}_0 {{\partial W}\over {\partial
I}} [I(\xi, I_0, 
\theta_0)] + \varepsilon \int^{2\pi/2\mu_0}_0 {{\partial H^{(1)}}\over
{\partial I}} [\theta (\xi, I_0, \theta_0), I(\xi, I_0, \theta_0),
2\mu_0 \xi]\, d\xi. \eqno (3.17b)$$
The integrated terms are smooth functions of $I_0$, $\theta_0$ since the
solutions depend smoothly on the initial values $(I_0, \theta_0)$ and 
the integration interval is finite.  Thus the map takes the form
$$I^{(1)} = I_0 + \varepsilon F(\theta_0, I_0), \eqno (3.18a)$$
$$\theta^{(1)} = \theta_0 + {{\pi}\over {\mu_0}} \omega (I_0) +
\varepsilon G(\theta_0, I_0), \eqno (3.18b)$$
where $\omega (I_0) \equiv \partial W/\partial I |_{I = I_0}$.

Since $I$ can be expanded in a power series in 
$\varepsilon$ (the function
is analytic in $\varepsilon$ because the equation is integrated over a finite
interval).  The integrated terms of $\omega (I)$ contribute to the term $G$
to order $\varepsilon$.  The contribution to the leading order just comes 
from $I_0$.

This is the canonical form of the map for the application of the K.A.M.
theorem [10].  The map for 
$\varepsilon = 0$ has as invariant curves the circles $I =$ constant,
$0 \leq \theta \leq 2\pi$.  Moreover on each circle the rate of 
advance is $\pi \omega (I)]/\mu_0$ which is action dependent.  (In this
simple case the invariant curves of the map are just the level lines of
the unperturbed Hamiltonian.)  In this setting the K.A.M. theorem
guarantees that for $E$ sufficiently small and $\omega (I)$ sufficiently
irrational the invariant curves persist [10].  It is also known that this set
of ``sufficiently irrational'' numbers has full measure.  Thus, since
the function $\omega (I)$ is increasing (and tends to infinity as
$I \rightarrow \infty$) it is clear that there is a set of full measure
in the variable $I$ for which invariant curves persist.  It then follows
that the motion of $a(t)$, $\dot a(t)$ is bounded for all times, provided
that  the initial conditions fall within the perturbed invariant tori.
This gives a rigorous verification of the perturbation results obtained
in the previous section.  

Up to this point we have made no reference to field theory.  The
results in this section are actually a convenient way to study
approximate coherent states in one-dimensional quantum mechanics. 
They can be applied to the motion of coherent states with
``breathing'' or ``spreading'' such as those studied by Guth and Pi
[11].
There are a number of ways to extend our results to field theory,
some of which will be mentioned briefly in Sec. V.
As mentioned in the Introduction,
one of the main objectives of this article was to use the variational
approach to investigate how close a microsuperspace solution is to a
larger minisuperspace solution over a long period of time.  In order
to do this, in the next section we will apply the methods used above
to the same minisuperspace-microsuperspace problem used in Ref. [4].
\vfill\eject

\noindent {\bf IV. MINISUPERSPACE FIELD THEORY}
\vskip 10 pt

In Ref. [4] coherent states
peaked around some small sector of a field theory
were used to investigate whether they
could be approximated by a state restricted to that sector and then
quantized.  As discussed in the Introduction, the system
was a $\lambda \varphi^4$ model in an $S^1$ topology with
only two modes ``unfrozen''.  That is, if in (1.1)
we put all but $\varphi_0$ and {\it one\/} of the $\varphi_n$ equal
to zero, we get a minisuperspace field theory by plugging this ansatz
into the $\lambda \varphi^4$ action.  Putting the
$\varphi_{-n}$ equal to zero and dropping third and fourth order
cross terms among the $\varphi_n$ in the action
yielded a two-dimensional classical theory that could be
quantized.  Moreover, using the definitions $\varphi_0 \equiv x$ and
$\varphi_n \equiv y$ it was shown that the minisuperspace state function
obeyed the Schr\"odinger equation
$$-{{1}\over {2}}{{\partial^2 \Psi}\over {\partial x^2}} - {{1}\over
{2}}{{\partial^2 \Psi}\over {\partial y^2}} + {{\mu_0^2}\over {2}}
x^2 \Psi + \varepsilon x^4 \Psi + {{m_0^2}\over {2}} y^2 \Psi +
6\varepsilon x^2 y^2 \Psi = i{{\partial \Psi}\over {\partial t}},
\eqno (4.1)$$
where $\varepsilon = \lambda/L$ and $m_0^2 = \mu_0^2 + (2\pi n/L)^2$.  In
Ref. [4] a pointwise approximation for $\Psi = e^{-\cal S}$ with ${\cal
S}$ similar to (2.12) was used.  Here we will use the variational
approach with $\Psi = W(t)e^{-\cal S}$, where
$${\cal S} = {{\mu^2}\over {2}} (x - g_1)^2 + {{m^2}\over {2}} (y -
g_2)^2 + \theta xy + iP_1 x + iP_2 y + iC_1 x^2 + iC_2 y^2 + iMxy + i\phi (t),
\eqno (4.2)$$
and $\mu$, $m$, $g_1$, $g_2$, $\theta$, $P_1$, $P_2$, $C_1$, $C_2$,
$M$ are all functions of $t$.  If we insert this ${\cal S}$ into
$$L = \int_{t_0}^{t_1}dt \int^{\infty}_{-\infty} dx
\int^{\infty}_{-\infty} dy [i\Psi^{*} \Psi_t -{{1}\over {2}}
\Psi^{*}_x \Psi_x - {{1}\over {2}} \Psi^{*}_y \Psi_y - V(x, y)
\Psi^{*} \Psi ], \eqno (4.3)$$
with 
$$V(x, y) = {{\mu_0^2}\over {2}} x^2 + \varepsilon x^4 + {{m_0^2}\over
{2}} y^2 + 6\varepsilon x^2 y^2, \eqno (4.4)$$
we find a very large action for $\mu$, $m$, $g_1$, $g_2$, $\theta$,
$P_1$, $P_2$, $C_1$, $C_2$, and $M$ which is given in Appendix B. Of
course, we are only interested in states that can be approximated by
microsuperspace states with $y = 0$, i.e. those with $g_2 = 0$.  If we
now look at the form of $\Psi^{*}\Psi$, we see that a $\Psi^{*}\Psi$ =
const. surface is an ellipse centered on $y = 0$, $x = g_1$ with its
semi-major and semi-minor axes given by $\mu$ and $m$ (which axis is
associated with which parameter depends on their relative size).  The
parameter $\theta$ gives the angle that this ellipse makes with the
$x$-axis.  There is some freedom in defining what one means by a
minisuperspace state ``close'' to a microsuperspace state (here the
$y = 0$ state).  For simplicity we would like to take $\theta = 0$
and $g_2 = 0$.  If one varies the action (B1) with respect to $g_2$,
$P_2$, $\theta$, and $M$, one finds a series of equations  which, for
the initial conditions $\dot g_2 = g_2 = 0$, $\dot P_2 = P_2 = 0$,
$\dot \theta = \theta = 0$, $\dot M = M = 0$, maintain $\theta$,
$g_2$, $P_2$, and $M$ zero for all time.  For this solution the
reduced action
$$L = \int_{t_0}^{t_1} {{\pi W^2}\over {\sqrt{m\mu}}} \bigg [ \dot P_1
g_1 + \dot C_1 \left (g_1^2 + {{1}\over {2\mu}} \right ) + {{\dot
C_2}\over {2m}} + \dot \varphi -$$
$$-\bigg \{ {{P_1^2}\over {2}} + {{\mu_0^2}\over {2}} g^2_1 +
\varepsilon g_1^4 + {{3\varepsilon g_1^2}\over {\mu}} + 2P_1 C_1 g_1
+$$
$$+2C_1^2 \left (g_1^2 + {{1}\over {2\mu}} \right ) + {{C_2^2}\over
{m}} + {{\mu}\over {4}} + {{\mu_0^2}\over {4\mu}} +
{{3\varepsilon}\over {4\mu^2}} + {{m}\over {4}} +$$
$$+ {{m_0^2}\over {4m}} + {{3\varepsilon}\over {m}} \left (g_1^2 +
{{1}\over {2\mu}} \right ) \bigg \} \bigg ] dt \eqno (4.5)$$
gives the correct equations of motion.

The Euler equations for this action are
$$-{{2C_2}\over {m}} + {{\dot m}\over {2m^2}} = 0, \eqno (4.6a)$$
$$-{{\dot C_2}\over {2m^2}} + {{C_2^2}\over {m^2}} - {{1}\over {4}} +
{{m_0^2}\over {4m^2}} + {{3\varepsilon}\over {m^2}} \left ( g_1^2 +
{{1}\over {2\mu}} \right ) = 0, \eqno (4.6b)$$
$$-\dot g_1 - P_1 - 2C_1 g_1 = 0, \eqno (4.6c)$$
$$\dot P_1 + 2g_1 \dot C_1 - \mu_0^2 g_1 - 4\varepsilon g_1^3 -
{{6\varepsilon g_1}\over {\mu}} - 2P_1 C_1 = 0, \eqno (4.6d)$$
$$-4C_1^2 g_1 - {{6\varepsilon}\over {m}} g_1 = 0, \eqno (4.6e)$$
$$-2g_1 \dot g_1 + {{1}\over {2}} {{\dot \mu}\over {\mu^2}} -2P_1 g_1
- 4C_1 \left ( g_1^2 + {{1}\over {2\mu}}\right ) = 0, \eqno (4.6f)$$
$$-{{\dot C_1^2}\over {2\mu^2}} + {{C_1^2}\over {\mu^2}} -
{{\mu_0^2}\over {4\mu^2}} + {{3\varepsilon g_1^2}\over {\mu^2}} +
{{3\varepsilon}\over {2\mu^2}} + {{3\varepsilon}\over {2m}} {{1}\over
{\mu^2}} = 0, \eqno (4.6g)$$
together with the normalization condition 
$(W^2/\sqrt{m\mu})^{\bf \cdot}
 = 0$ and an
equation for the trivial phase $\phi$. These equations give
$$-\ddot g_1 - \mu_0^2 g_1 - 4\varepsilon g^3_1 - {{6\varepsilon
g_1}\over {\mu}} - {{6\varepsilon g_1}\over {m}} = 0, \eqno (4.7a)$$
$$-{{\ddot \mu}\over {\mu}} + {{3}\over {2}} \left ( {{\dot \mu}\over
{\mu}} \right )^2 - 2\mu^2 + 2\mu_0^2 + 24 \varepsilon g_1^2 +
{{12\varepsilon}\over {m}} + {{12\varepsilon}\over {\mu}} = 0, \eqno
(4.7b)$$
$$-{{\ddot m}\over {m}} + {{3}\over {2}} \left ( {{\dot m}\over {m}}
\right )^2 - 2m^2 + 2m_0^2 + 24\varepsilon \left ( g_1^2 + {{1}\over
{2\mu}} \right ) = 0. \eqno (4.7c)$$

In this case we can also appeal to a K.A.M. type of result in order
to obtain qualitative information about the nature of the solutions. 
For this we observe that the unperturbed system
$$\ddot g_1 + \mu_0^2 g_1 = 0, \eqno (4.8a)$$
$$\ddot a + \mu_0^2 a - {{1}\over {a^3}} = 0, \eqno (4.8b)$$
$$\ddot b + \mu_0^2 b - {{1}\over {b^3}} = 0, \eqno (4.8c)$$
can be written in terms of action-angle variables.  The equations for
$a$ and $b$ have the variables given in (3.13) and (3.15).  The
equation for $g_1$ has the polar coordinates as action-angle
variables.  Denoting the actions by $I_g$, $I_a$, $I_b$ and the
angles by $\theta_g$, $\theta_a$, $\theta_b$, the perturbed problem
takes the form
$$\dot I_g = -\varepsilon {{\partial H^{(1)}}\over {\partial \theta_g}} \qquad
\dot \theta_g = \mu_0 + \varepsilon {{\partial H^{(1)}}\over
{\partial I_g}}, \eqno (4.9a)$$
$$\dot I_a = -\varepsilon {{\partial H^{(1)}}\over {\partial \theta_a}} \qquad
\dot \theta_a = \omega_a (I_a) + \varepsilon {{\partial H^{(1)}}\over
{\partial I_a}}, \eqno (4.9b)$$
$$\dot I_b = -\varepsilon {{\partial H^{(1)}}\over {\partial \theta_b}} \qquad
\dot \theta_b = \omega_b (I_b) + \varepsilon {{\partial H^{(1)}}\over
{\partial I_b}}, \eqno (4.9c)$$
Since the frequency $\mu_0$ does not depend on the actions, the usual
K.A.M. theorem cannot be applied to guarantee the persistence of the
invariant tori $I_g$ = const., $I_a$ = const., $I_b$ = const.,
instead a modification due to Melnikov, Poesch and Kuskin [8] can be
used.  The result is as follows.  If the unperturbed Hamiltonian
system is of the form 
$$\dot {\bf \Theta} = {\bf \Lambda}, \qquad \dot {\bf I} = 0, \eqno
(4.10)$$
where ${\bf \Lambda}$ is a constant vector, then small Hamiltonian
perturbations preserve the invariant tori ${\bf I}$ = const. ${\bf
\Theta} = {\bf \Lambda} t + {\bf \Phi}$ for most values of the
vector ${\bf \Lambda}$.  In this case we choose
$${\bf \Lambda} = [\mu_0, \omega_a (I_a^0 ), \omega_b (I^0_b )],
\eqno (4.11)$$
and thus the remainder is small provided that we search for tori
close to $I^0_a$, $I^0_b$.  The result guarantees that for most
values (a set of full measure) of $\mu_0$, $I^0_a$, $I^0_b$ the
unperturbed tori persist.

This shows the existence of quasiperiodic, and thus bounded, motions
for most initial conditions $I^0_a$, $I^0_b$.  In this case the
existence of invariant tori does not prove stability because three
dimensional tori do not separate the six-dimensional phase space. 
Moreover, a universal instabilty, known as Arnold diffusion is
present.  This implies that for sufficiently long times, that is,  
$t_{AD} = {\cal O}
(e^{1/\varepsilon})$ particular initial values always leave any
bounded region of phase space.  In general this time is very long
compared to other time scales in the problem. In order to see whether this
instability is relevant for our problem, it is necessary to rewrite
our equations in terms of more conventional units.  The expression
for the Arnold diffusion time mentioned above is actually $t_{AD} =
(1/\omega_0) e^{1/\epsilon}$, where $\epsilon$ is a constant
which estimates the ratio of a perturbation to a Hamiltonian to the
unperturbed Hamiltonian, and $\omega_0$ is the frequency of
oscillation associated with the unperturbed Hamiltonian.

The units of $\varphi$ in (1.2) are $Q\ell$, where $\ell$ is the unit
of length and $Q$ is a ``charge'' associated with $\varphi$.
The fact that we are working in one space dimension rather than three
means that $Q^2$ has units of force. The constant $\mu_0$ has units
of one over length, and
the constant $\lambda$ must have
units of the reciprocal of energy times length cubed, and
$\varepsilon$ has units of one over energy times length to the fourth
power.  This, along with the fact that $\partial/\partial t$ is
really $\partial/\partial ct$, allows us to determine the units of
$g$ and $a$ which are both $Q\ell^{3/2}$.  We can define the
dimensionless
variables, $u = g/QL^{3/2}$ and $b = a/QL^{3/2}$, and Eqs. (2.8)
become
$$\ddot u + c^2\mu_0^2u + 4\varepsilon_0 \left ( {{c}\over {L}}
\right )^2 u^2 + 6\varepsilon_0 \left ( {{c}\over {L}} \right )^2 b^2
u = 0, \eqno (4.12a)$$
$$\ddot b + c^2 \mu_0^2 b - \left ( {{c^4 \hbar^2}\over {Q^4 L^6}}
\right ) {{1}\over {b^3}} - 12 \varepsilon_0 \left ({{c}\over {L}}
\right )^2 b^3 + 24 \varepsilon_0 \left ( {{c}\over {L}} \right )^2
bu^2 = 0, \eqno (4.12b)$$
where $\varepsilon_0$ is a small dimensionless parameter.  By adjusting
$Q$ we can put $\varepsilon_0$ equal to one.   

We need only give $\mu_0$ and $Q$ to find $t_{AD}$.  These quantities
depend on the type of field that $\varphi$ is supposed to model.
Since $\varphi$ is one-dimensional, these quantities are somewhat
different from those one would expect in the three-dimensional case. 
For $\mu_0$, a classical quantity that has the proper units is 
$mc^2/e^2$, where we can take $e$ and $m$ the charge and the mass of
the proton, 
which gives $\omega_0 = mc^3/e^2$.  A quantity $Q$
associated with the charge that has the proper units is $Q = mc^2/e$
(note that the dimensionality forces us to take a reciprocal of $e$). 
This implies that $c^4 \hbar^2/Q^4 L^6$ is
$(1/\alpha_f^2)(\lambda_c/L)^2 (c/L)^2$, where $\lambda_c$ is the
Compton wavelength of the proton and $\alpha_f$ is the fine
structure constant.  This quantity is small with
respect to $c/L$ for $L > \lambda_c$, so $H_1$ is proportional to
$(c/L)^2$, while $H_0$ is proportional to $m^2 c^6/e^4$, so $t_{AD}$
is roughly $t_{AD} = (e^2/mc^3) \exp [(Lmc^2/e^2)^2]$.  In principle,
Arnold diffusion could be important when $L < e^2/mc^2$.  However, we
can take a cosmological model which, for simplicity, has the time
behavior of a $k = 0$ Robertson-Walker universe, but the mass of a $k
= 1$ model, which gives the age of the universe for any $L$ to be
$$t_u = {{L^{3/2}}\over {\sqrt {6\pi GM_0}}},$$  
where $M_0 = 5 \times 10^{56} {\rm g}$.

If we now take $L =
\beta (t) (e^2/mc^2)$, then the ratio of $t_{AD}$ to the age of the
universe at any time would be
$${{t_{AD}}\over {t_u}} = {{1}\over {\beta^{3/2}}} e^{\beta^2} \left
({{6\pi G M_0}\over {e^2/m}} \right )^{1/2} = {{1}\over {\beta^{3/2}}}
e^{\beta^2} (3.5 \times 10^{20}).$$
This quantity has a minimum of order one at $\beta$ of order one and
grows for all other values of $\beta$, so
$t_{AD}$ is always much larger that the age of the universe.

If $\varphi$ is supposed to model the gravitational field itself, we
can take $Q^2 = c^4/G$ (again the reciprocal of the usual
``charge'').  Here $c^4 \hbar^2 /Q^4 L^6$ is $(L_p /L)^4 (c/L)^2$,
where $L_p$ is the Planck length.  Since the field is massless, we
can take $\mu_0 = 0$.  Here, then, the $(c/L)^2$ terms are $H_0$, and
the term in $c^4 \hbar^2 /Q^4 L^6$ is $H_1$, so $\epsilon$ is of
order $(L_p /L)^4$, so $t_{AD}$ is roughly $t_{AD} = (L/c) \exp
[(L/L_p)^4]$, and the ratio of $t_{AD}$ to the age of the universe
(the age of the universe does not 
change) is (now taking $L = \beta (t) L_p$)
$$t_{AD} = {{1}\over {\sqrt {\beta}}} e^{\beta^4} \left ( {{6\pi G
M_0}\over {L_p c^2}}\right )^{1/2} = 
{{1}\over {\sqrt {\beta}}} e^{\beta^4} (7
\times 10^{31}),$$
which is, again, always much larger than one.

We thus conclude that the approximation based on coherent states is
good in the sense that classical trajectories of the center of mass
are not qualitatively modified by the existence of small numbers of
higher modes in field theory.  However, when infinitely many modes
are present it is not possible to draw any rigorous conclusions about
the limit.  This case has to be investigated independently.   Even
simple finite-mode systems exhibit new features in the limit.
\vfill\eject  
 
\noindent {\bf V. CONCLUSIONS AND SUGGESTIONS FOR FURTHER RESEARCH}
\vskip 10 pt

The twofold purpose of this paper has been addressed by means of the
approximation techniques given in Section II.
We have shown that it is possible to obtain time-dependent consistent 
approximations to the motion and breathing of coherent states using
average Lagrangians. These Lagrangians have allowed us to explore not
only the evolution of these states in quantum mechanics, but have
allowed us to study the stability of such solutions in minisuperspace
field theory.  
The most severe potential instability arose from
the resonant interaction of the center of mass of the state and the
width parameter.  We showed, using a technique of 
averaging of ordinary differential
equations, how nonlinear effects saturate the resonance amplitude
and change the solution into a modulated solution.  In special cases
the stability results can be proved rigorously by means of the K.A.M.
theorem.  This formalism suggests a new potential instability when
many modes of comparable size are present in the coherent state.
In fact, a finite but high dimensional Hamiltonian system is obtained.  In
this case Arnold diffusion could have a destabilizing effect over very
long times.  Clearly the same instability will be present in any
Hamiltonian truncation of more realistic field theoretic models, and its
relevance in practice will have to be studied in each case.

Finally, we would like to emphasize the fact that this variational idea 
could be applied to functional Lagrangians.
The parameters will now be functions of both position and time.  The
averaged action will have as its Euler equations nonlinear partial
differential equations which will describe the spread and the position
of the coherent state in function space.  This is currently under
investigation for a simple $\varphi^4$ functional model.
\vfill\eject   

\noindent {\bf APPENDIX A. POINTWISE APPROXIMATION EQUATIONS }
\vskip 10 pt

If we write $\Psi = e^{-{\cal S}}$ with ${\cal S}$ given by (2.12),
then ${\cal S}$ obeys
$$i{{\partial {\cal S}}\over {\partial t}} + {{1}\over {2}}
{{\partial^2 {\cal S}}\over {\partial x^2}} - {{1}\over {2}} \left (
{{\partial {\cal S}}\over {\partial x}} \right )^2 + V(x) = 0.$$
Inserting (2.12) in this equation and assuming that $\alpha^2$,
$\beta^2$ and $B^2$ are small enough to be ignored, we can equate the
real and imaginary coeeficients of each power of $x$ on the LHS equal
to zero.  We find
$$x^4: \qquad \qquad -4\alpha \mu + \varepsilon = 0 \qquad ({\rm real})
\eqno (A1)$$
$$ \dot \alpha - 8 \alpha C = 0 \qquad (Imag.) \eqno (A2)$$
$$x^3: \qquad 4\alpha \mu g - 3\beta \mu + 6BC - \dot B = 0 \qquad
({\rm Real}) \eqno (A3)$$
$$\dot \beta - 4\alpha P - 6C\beta - 3B \mu = 0 \qquad ({\rm Imag.})
\eqno (A4)$$
$$x^2: \qquad 6\alpha - \dot C - {{\mu^2}\over {2}} + 2C^2 + 3\beta
\mu g + 3P\beta + {{\mu_0^2}\over {2}} = 0 \qquad ({\rm Real}) \eqno
(A5)$$
$${{\dot \mu}\over {2}} - 3\beta P + 3B \mu g - 2\mu C = 0
\qquad ({\rm Imag.}) \eqno (A6)$$
$$x: \qquad 3\beta - \dot P + \mu^2 g + 2PC = 0 \qquad ({\rm Real})
\eqno (A7)$$
$$-\dot \mu g - \mu \dot g + 3B + 2\mu C g - \mu P = 0 \qquad
({\rm Imag.}) \eqno (A8)$$
\vfill\eject

\noindent {\bf APPENDIX B. COMPLETE TWO-DIMENSIONAL ACTION}
\vskip 10 pt

The action (4.3) for $\Psi = W(t) e^{-{\cal S}}$ with ${\cal S}$ given by
(4.2) is, after integration over $x$ and $y$,
$$\int^{t_1}_{t_0} {{\pi W^2 e^A}\over {\sqrt{m\mu - \theta^2}}}
\bigg [ \dot  P_1 \left ( g_1 - {{m\theta g_2 - \theta^2 g_1}\over
{m\mu - \theta^2}} \right ) + \dot P_2 \left ( g_2 - {{\mu \theta g_1
- \theta^2 g_2}\over {m\mu - \theta^2}} \right ) +$$
$$+ \dot C_1 \left ( g_1^2 + {{1}\over {2\mu}} - 2\theta g_1 \left (
{{mg_2 -\theta g_1}\over {m\mu - \theta^2}} \right ) + {{m\theta^2 -
\theta^4/\mu + 2(mg_2 - \theta g_1 )^2 \theta^2}\over {2(m\mu -
\theta^2)^2}} \right )+$$
$$+ \dot C_2 \left (g_2^2 + {{1}\over {2m}} - {{2\theta\mu g_1 g_2 -
2\theta^2 g_2^2}\over {m\mu - \theta^2}} + {{\mu \theta^2 -
\theta^4/m + 2(\mu g_1 - \theta g_2 )^2 \theta^2}\over {2(m\mu -
\theta^2 )^2}} \right ) +$$
$$+ \dot M\left ( {{\mu m g_1 g_2 - \mu \theta g_1^2}\over {m\mu
-\theta^2}} - {{\theta}\over {2(m\mu - \theta^2)}} - {{\mu \theta
(mg_2 - \theta g_1 )^2}\over {2(m\mu - \theta^2)^2}} \right ) -$$
$$ - {{1}\over {2}} \mu^2 \left ( {{1}\over {2\mu}} + {{m\theta^2 -
\theta^4/\mu + 2(mg_2 - \theta g_1)^2 \theta^2}\over {2(m \mu -
\theta^2)^2}} \right ) -$$
$$-\mu \theta \left ( {{\mu m g_1 g_2 - \mu \theta g_1^2}\over {m\mu
- \theta^2}} - {{\theta}\over {2(m\mu - \theta^2)}} - {{\mu\theta
(mg_2 - \theta g_1 )^2}\over {(m\mu - \theta ^2)^2}} - g_1 \left [
g_2 - {{\mu \theta g_1 - \theta^2 g_2}\over {m\mu - \theta^2}} \right
] \right ) -$$
$$- {{\theta^2}\over {2}} \left ( g_2^2 + {{1}\over {2m}} - {{2\mu
\theta g_1 g_2 - 2\theta^2 g_2^2}\over {m\mu - \theta^2}} + {{\mu
\theta^2 - \theta^4/m + 2(\mu g_1 - \theta g_2 )^2 \theta^2}\over
{2(m\mu - \theta^2 )^2}} \right ) -$$
$$- {{P_1^2}\over {2}} - 2C_1^2 \left ( g_1^2 + {{1}\over {2\mu}} -
{{2m\theta g_1 g_2 - 2\theta^2 g_1^2}\over {m\mu - \theta^2}} +
{{m\theta^2 - \theta^4/\mu + 2(mg_2 - \theta g_1)^2 \theta^2}\over
{2(mg_2 - \theta^2)^2}}\right ) -$$
$$- {{1}\over {2}} M^2 \left ( g_2^2 + {{1}\over {2m}} - {{2\mu
\theta g_1 g_2 - 2\theta^2 g_2^2}\over {m\mu - \theta^2}} + {{\mu
\theta^2 - \theta^4/m + 2(\mu g_1 - \theta g_2)^2 \theta^2}\over
{2(m\mu - \theta^2)^2}} \right ) -$$
$$-2P_1 C_1 \left (g_1 - {{m\theta g_2 - \theta^2 g_1}\over {m\mu -
\theta^2}}\right ) - P_1 M \left ( g_2 - {{m\theta g_1 - \theta^2
g_2}\over {m\mu - \theta^2}} \right ) -$$
$$-2C_1 M \left ({{m\mu g_1 g_2 - \mu \theta g_1^2}\over {m\mu -
\theta^2}} - {{\theta}\over {2(m\mu - \theta^2)^2}} - {{\mu \theta
(m g_2 - \theta g_1)^2}\over {(m\mu - \theta^2 )^2}} \right ) -$$
$$- {{m^2}\over {2}} \left ( {{1}\over {2m}} + {{\mu \theta^2 -
\theta^4 /m + 2(\mu g_1 - \theta g_2 )^2 \theta^2}\over {2(m\mu -
\theta^2)^2}}\right ) -$$
$$- m\theta \left [ {{m\mu g_1 g_2 - \mu \theta g_1^2}\over {m\mu -
\theta^2}} - {{\theta}\over {2(m\mu - \theta^2)}} - {{\mu \theta
(mg_2 - \theta g_1)^2}\over {(m\mu - \theta^2)^2}} - g_2\left (g_1 -
{{m\theta g_2 - \theta^2 g_1}\over {m\mu - \theta^2}} \right ) \right
] -$$
$$-{{\theta^2}\over {2}} \left ( g_1^2 + {{1}\over {2\mu}} -
{{2m\theta g_1 g_2 - 2\theta^2 g_1}\over {m\mu - \theta^2}} +
{{m\theta^2 - \theta^4/\mu + 2(mg_2 - \theta g_1)^2 \theta^2}\over
{2(m\mu - \theta^2)^2}} \right ) -$$
$$-{{1}\over {2}} P^2_2 - 2C_2^2 \left( g_2^2 + {{1}\over {2m}} -
{{2\mu \theta g_1 g_2 - 2\theta^2 g_2^2}\over {m\mu - \theta^2}} +
{{\mu \theta^2 - \theta^4/m + 2(\mu g_1 - \theta g_2)^2
\theta^2}\over {2(m \mu - \theta^2)^2}}\right ) -$$
$$-{{1}\over {2}} M^2\left ( g_1^2 + {{1}\over {2\mu}} - {{2m\theta
g_1 g_2 - 2\theta^2 g_1^2}\over {m\mu - \theta^2}} + {{m\theta^2 -
\theta^4/\mu + 2(mg_2 - \theta g_1)^2 \theta^2}\over {2(m\mu -
\theta^2)^2}} \right ) -$$
$$-2P_2 C_2 \left ( g_2 - {{\mu \theta g_1 - \theta^2 g_2}\over {m\mu
- \theta^2}} \right ) - P_2 M \left ( g_1 - {{m\theta g_2 - \theta^2
g_1}\over {m\mu - \theta^2}} \right ) -$$
$$-2C_2 M \left ( {{\mu m g_1 g_2 - \mu \theta g_1^2}\over {m\mu -
\theta^2}} - {{\theta}\over {2(m\mu - \theta^2)}} - {{\mu \theta
(mg_2 - \theta g_1)^2}\over {(m\mu - \theta^2)^2}} \right ) -$$
$$- {{\mu_0^2}\over {2}}\left (g_1^2 + {{1}\over {2\mu}} - {{2m\theta
g_1 g_2 - 2\theta^2 g_1^2}\over {m\mu - \theta^2}} + {{m\theta^2 -
\theta^4/\mu + @(mg_2 - \theta g_1)^2 \theta^2}\over {2(m\mu -
\theta^2)^2}} \right ) -$$
$$- \varepsilon \bigg [ g_1^4 + {{3g_1^2}\over {\mu}} + {{3}\over
{4\mu^2}} - \theta {{(4g_1^3 + 6g_1/\mu)(mg_2 - \theta g_1)}\over
{m\mu - \theta^2}} +$$
$$+{{3\theta^2}\over {(m\mu - \theta^2)^2}} \left ( 2\mu^2 g_1^2 +
{{1}\over {\mu}}\right ) \left ( {{1}\over {2}} [m\mu - \theta^2] +
[mg_2 - \theta g_1]^2 \right ) -$$
$$- 4\theta^3 g_1 \left ( {{[mg_2 - \theta g_1]^3}\over {[m\mu -
\theta^2]^3}} + {{3}\over {2\mu}} {{[mg_2 - \theta g_1]}\over {[m\mu -
\theta^2]^2}} \right ) +$$
$$+ \theta^4 \left ( {{[mg_2 - \theta g_1]^4}\over {[m\mu -
\theta^2]^4}} + {{3}\over {\mu}} {{[mg_2 - \theta g_1 ]^2}\over
{[m\mu - \theta^2]^3}} + {{3}\over {4\mu^2}} {{1}\over {[m\mu -
\theta^2]^2}} \right ) \bigg ] -$$
$$- {{m_0^2}\over {2}} \left ( g_2^2 + {{1}\over {2m}} - {{2\mu
\theta g_1 g_2 - 2\theta^2 g_2^2}\over {m\mu - \theta^2}} + {{\mu
\theta^2 - \theta^4/m + 2(\mu g_1 - \theta g_2 )^2 \theta^2}\over
{2(m\mu - \theta^2)^2}} \right ) -$$
$$-6\varepsilon \bigg [ \left ( {{1}\over {2\mu}} + g_1^2 \right )
\left ( {{\mu}\over {2(m\mu - \theta^2)}} + {{\mu^2 (mg_2 - \theta
g_1 )^2}\over {(m\mu - \theta^2 )^2 }} \right ) -$$
$$- {{2\theta}\over {\mu}} g_1 \left ( {{\mu^3[mg_2 - \theta
g_1]^3}\over {[m\mu - \theta^2]^3}} + {{3}\over {2}} {{\mu^2[mg_2 - \theta
g_1]}\over {[m\mu - \theta^2]^2}} \right ) +$$
$$+ {{\theta^2}\over {\mu^2}} \left ( {{\mu^4[mg_2 - \theta g_1]^4}\over
{[m\mu - \theta^2]^4}} + 3 {{\mu^3[ mg_2 - \theta g_1 ]^2}\over {[m\mu
- \theta^2]^3}} + {{3}\over {4}} {{\mu^2}\over {[m\mu -
\theta^2]^2}}\right )
\bigg ] \bigg ] dt,
\eqno (B1)$$  
where $A = (\theta^2 [\mu g_1^2 + mg^2_2] - 2m\mu g_1 g_2)/(m\mu -
\theta^2)$.
\vfill\eject

\noindent {\bf REFERENCES}
\vskip 10 pt

\noindent \item {1.} C. Misner, in {\it Magic without Magic: John Archibald 
Wheeler\/}, J. Klauder, Ed. (Freeman, San Francisco, 1972).
\vskip 10 pt

\noindent \item {2.} C. Misner, Phys. Rev D {\bf 8}, 3271 (1973).
\vskip 10 pt

\noindent \item {3.} K. Kucha\v r and M. Ryan, in {\it Gravitational
Collapse and Relativity\/}, H. Sato and T. Nakamura, Eds. (World
Scientific, Singapore, 1986).
\vskip 10 pt

\noindent \item {4.} M. Rosenbaum, M. Ryan, and S. Sinha, Phys. Rev. D
{\bf 47}, 4443 (1993).
\vskip 10 pt

\noindent \item {5.} G. Whitham, {\it Linear and Nonlinear Waves\/} (John
Wiley \& Sons, New York, 1974).
\vskip 10 pt

\noindent \item {6.} A. A. Minzoni, {\it A review of recent results
in the perturbation theory of solitary waves\/}, Boletin Sociedad
Matematica Mexicana, and references therein, To appear, Feb. 1997.
\vskip 10 pt

\noindent \item {7.} J. Guckenheimer and P. Holmes, {\it Nonlinear
Oscillations, Dynamical Systems and Bifurcation of Vector Fields\/}
(Springer-Verlag, Heidelberg, 1983).
\vskip 10 pt

\noindent \item {8.} S. B. Kuskin, {\it Nearly Integrable Infinite
Dimensional Hamiltonian Systems\/} (Springer-Verlag, Heidelberg,
1991).
\vskip 10 pt

\noindent \item {9.} J. Gaveia and A. A, Minzoni, Contemporary
Mathematics {\bf 58}, Part III, 71 (1987).
\vskip 10 pt

\noindent \item {10.} J. Moser, {\it Stable and Random Motions of
Hamiltonian Systems\/} (Princeton Mathematical Studies) (Princeton U.
Press, Princeton, 1980).
\vskip 10 pt

\noindent \item {11.} A. Guth and S. Y. Pi, Phys. Rev. D {\bf 32},
1899 (1985). 
\end